\documentclass[referee]{raa}           
\usepackage{graphicx,times}
\usepackage{natbib}
\usepackage{amssymb,amsmath}
\bibpunct{(}{)}{;}{a}{}{,}

\usepackage[a4paper=true,pagebackref=true]{hyperref}
\hypersetup{pdftitle = The title of my PDF, pdfauthor = My name, pdfsubject= The subject, pdfkeywords = keyword1 keyword2 keyword3} 
\hypersetup{colorlinks = true, linkcolor = green, anchorcolor = red, citecolor = blue, filecolor = red, pagecolor = red, urlcolor = red}

\usepackage{float}

\begin{document}

   \title{Deformation measurement by single spherical near-field intensity measurement for large reflector antenna

\footnotetext{\small $*$ Supported by the National Natural Science Foundation of China.}
}

 \volnopage{ {\bf 20XX} Vol.\ {\bf X} No. {\bf XX}, 000--000}
   \setcounter{page}{1}

   \author{Qian Ye\inst{1,2}, Boyang Wang\inst{2}, Qiang Yao\inst{2}, Jinqing Wang
      \inst{1},  Qinghui Liu\inst{1}, Zhiqiang Shen\inst{1}}

   \institute{ Shanghai Astronomical Observatory, Chinese Academy of Sciences, Shanghai 200030, 
China; {\it yeqian@sjtu.edu.cn ,  jqwang@shao.ac.cn, liuqh@shao.ac.cn, zshen@shao.ac.cn}\\
        \and
             School of Mechanical Engineering, Shanghai Jiaotong Unversity,
             Shanghai 200240, China; {\it wby920422@sjtu.edu.cn }\\
\vs \no
   {\small Received 20XX Month Day; accepted 20XX Month Day}
}

\abstract{This paper presents a new method to obtain the deformation distribution on the main reflector of an antenna only by measuring the electric intensity on a spherical surface with the focal point as the center of the sphere, regardless of phase. 
Combining the differential geometry theory with geometric optics method, this paper has derived a deformation-intensity equation to relate the surface deformation to the intensity distribution of a spherical near-field directly. Based on the Finite difference method (FDM) and Gauss-Seidel iteration, deformation has been calculated from intensity simulated by GO and PO method, respectively, with relatively small errors, which prove the effectiveness of the equation proposed in this paper. 
By means of this method , it is possible to measure the deformation only by scanning the electric intensity of a single hemispherical near-field whose area is only about $1/15$ of the aperture. And the measurement only needs a plane wave at any frequency as the incident wave, which means that both the signals from the outer space satellite and the far-field artificial beacon could be used as the sources. The scanning can be realized no matter what attitude and elevation angle the antenna is in because the size and angle of the hemisphere are changeable.
\keywords{telescopes --- waves --- line: profiles --- scattering 
}
}

   \authorrunning{Q. Ye, BY. Wang, Q. Yao, et al. }            
   \titlerunning{Deformation measurement by a hemisphere scanning}  
   \maketitle

%
\section{Introduction}           
\label{sect:intro}

The surface deformation of the antenna seriously affects its performance and observation high frequency efficiency of the antenna(\citealt{Ruze+1966}). With the requirement of today’s real-time surface adjustment of large-diameter antenna , lots of high precision actuators have been employed under many main-reflectors of antennas to compensate for surface deformation due to its weight(\citealt{Wang+etal+2014}), temperature(\citealt{Huang+etal+2016}) or wind(\citealt{Zhang+etal+2015}). However, high-accuracy and real-time deformation measurement is necessary before the compensation.

It has been recognized that radio holography method is a powerful tool to measure the deformation of the reflector surface(\citealt{Rahmat-Samii+1984, Greve+Morris+2005}). The traditional theoretical basis of the radio holography methods including phase coherent and phase retrieval holography is the Fourier transform relation between the far-field pattern and the aperture field distribution(\citealt{Goodman+2005}). The phase coherent method needs to measure the intensity and phase distribution of the far-field simultaneously. Therefor it needs to set another antenna and a dual channel reference receiver to provide the reference phase. Although the measurement accuracy of this method can reach $25$ microns when expressed as axially resolved surface errors, it must pause at each measuring point to complete the relevant calculation, which will inevitably lead to a more complex process(\citealt{Morris+etal+1988}). And it is not applicable in all frequencies and elevation angles of antenna due to the limitation of strong signal source with high frequency.

Phase retrieval holography like Misell method does not need the phase distribution of the far-field, and therefore no more equipment is required(\citealt{Morris+1996}). But it needs to measure the intensity of the far-field patterns for several times with different de-focus values or different phase shifts, resulting in a longer measuring time. The root-mean-square (RMS) measurement errors of the phase retrieval method in the derived aperture plane phase distribution can reach about $50$ microns while the original specification for the surface accuracy called for a RMS error of $100$ microns(\citealt{Morris+etal+2009}). Meanwhile, a high signal-to-noise ratio (SNR) to meet the requirement of measurement is extremely hard to realize(\citealt{Yaccarino+Rahmat-Samii+1997}). Therefore, this method has only a few previous astronomical applications. Out of focus (OOF) holography requires a smaller dynamic range, which can make use of astronomical sources and receivers. Besides, it works at full elevation angle and a relatively lower SNR condition, making OOF holography method successfully applied to many antennas. However, its poor signal-to-noise ratio and resolution lead to a low accuracy of measurement. The random error associated with OOF technique of the most famous application on the 100-m Green Bank Telescope (GBT) is about $\lambda / 100$, where $\lambda$ is the wavelength of the signal(\citealt{Nikolic+etal+2007}). Moreover, atmospheric disturbance caused by bad weather and the drastic change of visibility will make the two methods mentioned above unusable.

The traditional methodology of near-field scanning for small antennas requires a scanning frame with a probe on it. The scanning of the reflector can be realized by moving the frame along $x, y$ and $z$ directions, to obtain the near-field intensity (or phase) distributions of antennas, typically, on a near-field planar grid. The SNR of this method is usually high enough and the near-field intensity and phase of signals can be obtained simultaneously by a vector network analyzer (VNA), while with the aperture of the antenna becomes larger, the scanning area increases exponentially, which results in the doubling of working time(\citealt{Ghaffar+etal+2008, Wang+etal+2020, Morris+2007}). At the same time, the scanning is hard to implement on a heavy aperture reflector due to the huge equipment used by this method. A Successful measurement with the near-field holography technique has been reported for Atacama Large Millimeter Array (ALMA), using a monochromatic transmitter at a frequency of $104.02$ GHz and located on a 50-m-high tower at a distance of only $315$ m from the antennas. It finally achieved the goal to have a surface accuracy of $25 \mu \mathrm{m}$ RMS while the original surface accuracy is no more than $100 \mu \mathrm{m}$. However, the elevation angle was approximately $9^{\circ}$ unchanged, which means that this method can never work at full elevation angle once the transmitter is fixed(\citealt{Baars+etal+2007}).

The high measurement accuracy of all the methods mentioned above have obvious dependence on the high frequency of the transmitter, however, due to the lack of the EHF transmitter such as in Q-band($30-50$GHZ) or even W-band($75-110$GHZ), these methods are hard to provide a measure technology of high precision in all weathers and all attitudes. \cite{Huang+etal+2017} from Shanghai Jiao Tong university has presented a near-field measurement method by scanning a planar grid, which is insensitive to position error and independent of frequency. However, it still has some disadvantages such as the large scanning area, difficulties of implementation when the elevation angle changes and the need for additional mobile vehicles, which make it not suitable for large antennas. Moreover, the antenna is in the transmitting state when using this method to measure the deformation, which is contrary to the normal working state and will increase the complexity of the feed.

Therefore, the existing methods can’t satisfy the requirements of all-weather, all elevation, high SNR, high accuracy, frequency independent and quasi real-time measurement for the large-diameter antenna at the same time. In addition, for a dual-reflector antenna, the calculated deformation includes errors from the main reflector and the sub reflector together by the methods mentioned above, which are hard to be separated precisely(\citealt{Yu+etal+2015}).
\begin{table}[H]
   \centering
   \caption{Comparison of measurement methods for large-diameter antenna surface }
   \label{Tab1}
   \includegraphics[width=\textwidth, angle=0]{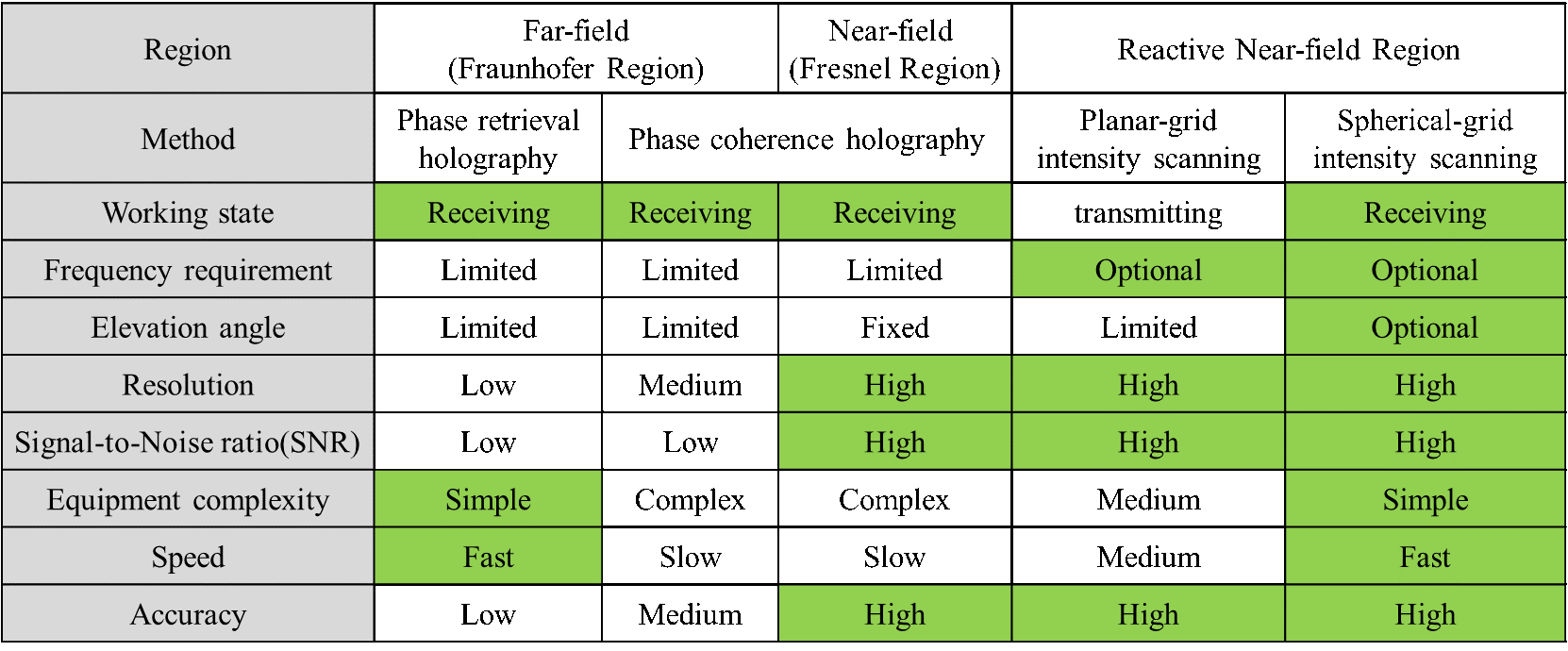}
   \end{table}
This paper presents a method which can measure the deformation of the main reflector independently, and its scanning area is only about $1/15$ of the aperture, which means that the scanning time will be greatly shortened. And this method can recover the deformation only by single electric intensity measurement regardless of phase, which needs no more equipment. Moreover, the measurement can be applied whenever the antenna is receiving signals at any frequency so that it won’t be limited by the attitude or the elevation angle and the feed, and both the signals from the outer space satellite and the far-field artificial beacon are available to be the sources of measurement. Meanwhile, in conjunction with other surface recovery methods, it is also expected to solve the deformation on the sub-reflector.

In order to achieve the goals above, as is shown in Table~\ref{Tab1}, our best choice is to select the near-field scanning method with high SNR, which is suitable for the strong sources at any frequency. To improve the measuring speed, we expect to effectively reduce the scanning area. And in order to make it suitable for transmitters at any frequency, we need to use a frequency independent surface recovery algorithm and expand its application to using an artificial beacon or a far-field astronomical source keeping the antenna receiving state, which can be regarded as plane wave to the antenna because of paraxial condition(\citealt{Rahmat-Samii+1984, Goodman+2005}). 
   \begin{figure}[H]
   \centering
   \includegraphics[width=0.8\textwidth, angle=0]{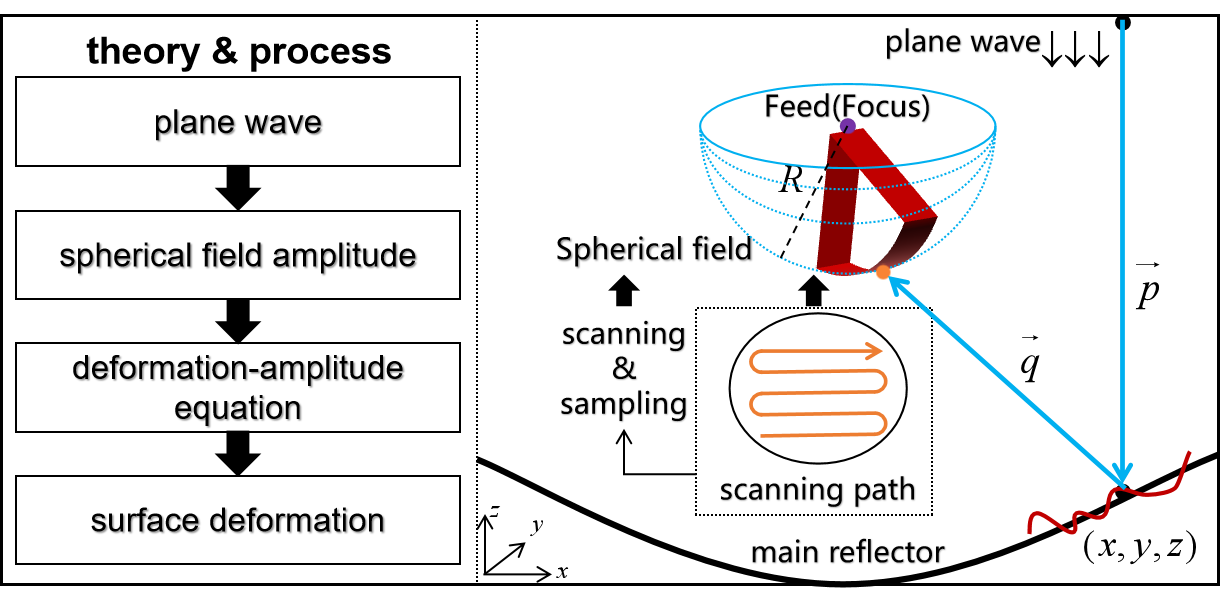}
   \caption{Schematic diagram of deformation recovery based on spherical field measurement}
   \label{Fig2}
   \end{figure}
This paper will discuss the feasibility of changing the original near-field planar grid to a hemispherical grid in the reactive near-field region with the second focus as the center to scan the intensity of the electromagnetic wave, as shown in Fig~\ref{Fig2}. In this process, we deduce the electric field complex amplitudes $E$ and $E^{*}$ reflected by the ideal antenna surface and the deformed one, respectively, with the plane wave as the radiation source. Then we find out the relation between the spherical field intensity and the deformation, which helps us to obtain and recover the deformation on the main reflector of the antenna from a relatively smaller scanning area.
   \begin{figure}[H]
   \centering
   \includegraphics[width=0.6\textwidth, angle=0]{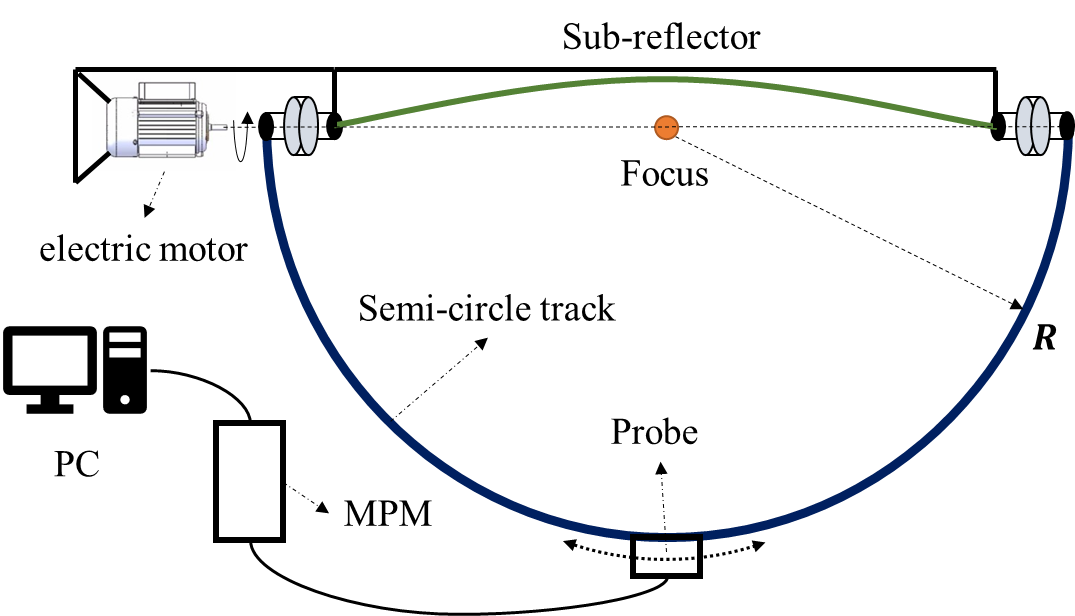}
   \caption{Schematic diagram of the implementation of the amplitude scanning}
   \label{Fig111}
   \end{figure}
An implementation method of the amplitude scanning on the hemispherical surface is described in Fig~\ref{Fig111}. The blue line denotes a semi-circle track with a radius of R, whose center is the focal point of the main reflector. The probe is fixed on the slider and moves along the track driven by a motor. Meanwhile, the track could rotate around the shaft fixed below the sub-reflector, which means that the sampling points can cover the whole hemispherical surface. The signal received by the probe is transmitted to the microwave power meter (MPM) through the test cable, and then we can obtain the amplitude data which the algorithm needs from the PC. In addition, when the antenna is in working state, the whole semi-circle track could be controlled to rotate to the position over the sub-reflector, so that all hardware facilities of the system will not interfere with the optical path of the antenna. And it is obvious that the scanning can be realized when the antenna is in any elevation angle.

\section{Mathematics}     
\label{sect:Math}
\subsection{Curvature Matrix of the Main Reflector} 
Let it first be assumed that the surface irregularities are described by function $\delta(r, \varphi)$ in the normal direction, and the main reflector can be expressed as a rotating paraboloid $z(x, y)$ with deformation $\delta(r, \varphi)$ ,namely,
\begin{equation}f=z+\delta(r, \varphi)=\frac{r^{2}}{4 F}+\delta(r, \varphi).\end{equation}
Using the concept of the first and second fundamental form of a curve, let $Z=\vec{Z}(r \cos \varphi, r \sin \varphi, f(r, \varphi))$ be a point of the main parabolic reflector, the first fundamental form of which at Z can be expressed as(\citealt{Pressley+2010})
\begin{equation}E d r^{2}+2 F d r d \varphi+G d \varphi^{2},\end{equation}
where

\begin{equation}E=\left\langle\vec{Z_{r}}, \vec{Z_{r}}\right\rangle=1+f_{r}^{2}, F=\left\langle\vec{Z_{r}}, \vec{Z_{\varphi}}\right\rangle=f_{r} f_{\varphi}, G=\left\langle\vec{Z_{\varphi}}, \vec{Z_{\varphi}}\right\rangle=r^{2}+f_{\varphi}^{2}.\end{equation}

We suppose $f_{r}=\partial f / \partial r, f_{\varphi}=\partial f / \partial \varphi$. Note that coefficients $E,F,G$ in Eq.(3) are called the first fundamental quantities and Eq.(2) allows one to compute lengths on a surface, and also angles and areas.

And the second fundamental form(\citealt{Pressley+2010}) is
\begin{equation}\frac{1}{2}\left(L d r^{2}+2 M d r d \varphi+N d \varphi^{2}\right),\end{equation}
where
\begin{equation}
\begin{split}
\begin{aligned}
&L=\left\langle\vec{n}, \vec{Z_{r r}}\right\rangle=\frac{r f_{r r}}{\sqrt{r^{2} f_{r}^{2}+f_{\varphi}^{2}+r^{2}}}\\
&M=\left\langle\vec{n}, \vec{Z_{r \varphi}}\right\rangle=\frac{r f_{r \varphi}-f_{\varphi}{ }^{2}}{\sqrt{r^{2} f_{r}^{2}+f_{\varphi}^{2}+r^{2}}}\\
&N=\left\langle\vec{n}, \vec{Z_{\varphi \varphi}}\right\rangle=\frac{r^{2} f_{r}+r f_{\varphi \varphi}}{\sqrt{r^{2} f_{r}^{2}+f_{\varphi}^{2}+r^{2}}}.
\end{aligned}
\end{split}
\end{equation}

We suppose $f_{r r}=\partial^{2} f / \partial r^{2}, f_{r \varphi}=\partial^{2} f /(\partial r \partial \varphi), f_{\varphi \varphi}=\partial^{2} f / \partial \varphi^{2}$. Note that coefficients $L,M,N$ in Eq.(5) are called the second fundamental quantities and Eq.(4) characterizes the degree of bend. 

Additionally, the unit normal $\vec{n}$ at $Z$ can be obtained.
\begin{equation}\vec{n}=\frac{\overrightarrow{Z_{r}} \times \overrightarrow{Z_{\varphi}}}{\left|\overrightarrow{Z_{r}} \times \overrightarrow{Z_{\varphi}}\right|}=\frac{1}{\sqrt{r^{2} f_{r}^{2}+f_{\varphi}^{2}+r^{2}}}\left(f_{\varphi} \sin \varphi-f_{r} r \cos \varphi,-f_{\varphi} \cos \varphi-f_{r} r \sin \varphi, r\right). \end{equation}

Based on the well-known Weingarten maps(\citealt{Pressley+2010}), the curvature matrix of the main reflector at $Z$ can be given as
\begin{equation}W=\left(\begin{array}{ll}a_{11} & a_{12} \\ a_{21} & a_{22}\end{array}\right)=\frac{1}{E G-F^{2}}\left(\begin{array}{ll}L G-M F & M E-L F \\ M G-N F & N E-M F\end{array}\right). \end{equation}

\begin{figure}[htbp]
\centering
\subfloat[]{
\label{fig3a}
  \begin{minipage}[t]{0.33\textwidth}
  \centering
  \includegraphics[width=2in]{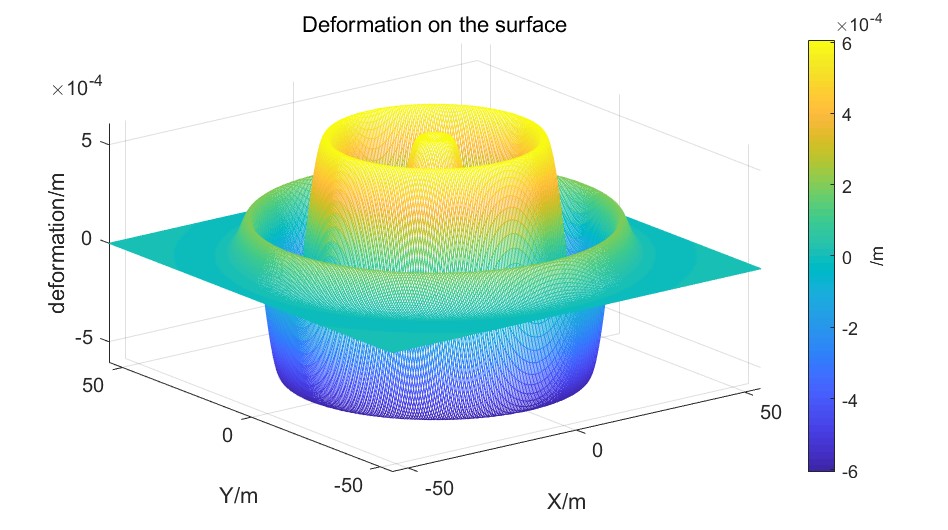}
  \end{minipage}
}
\subfloat[]{
\label{fig3b}
  \begin{minipage}[t]{0.33\textwidth}
  \centering
  \includegraphics[width=2in]{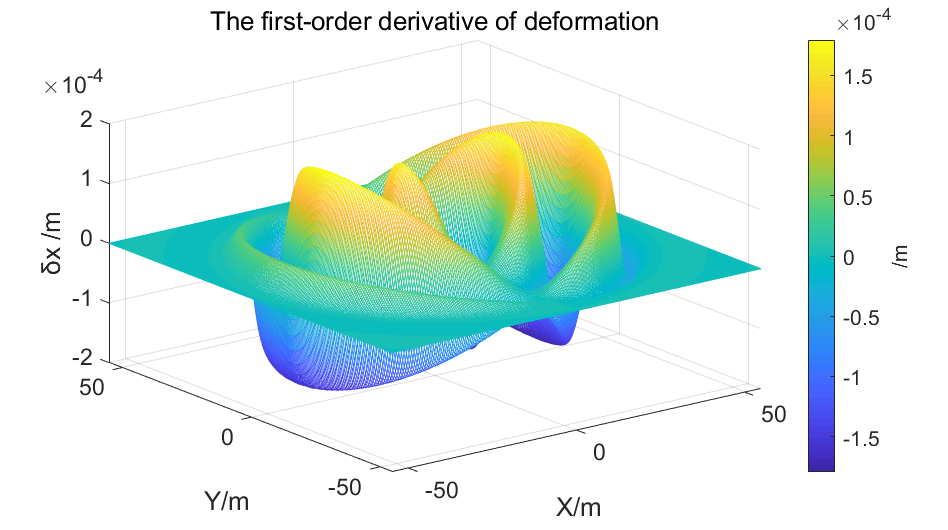}
  \end{minipage}
}
\subfloat[]{
\label{fig3c}
  \begin{minipage}[t]{0.33\textwidth}
  \centering
  \includegraphics[width=2in]{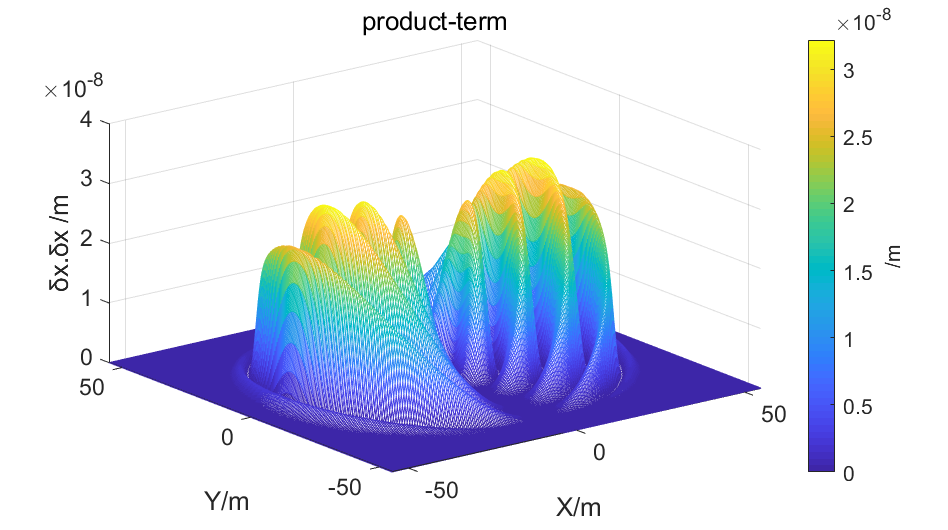}
  \end{minipage}
}
\caption{Magnitudes of deformation, the first partial derivative and the product order}
\end{figure}

Generally, the magnitude of deformation $\delta$ on the reflector is about $10^{-3} \mathrm{~m}$. Here we suppose a global smooth deformation on the surface as shown in Eq.(8) and Fig~\ref{fig3a}, with a magnitude of $1.2 \times 10^{-3} \mathrm{~m}$.
\begin{equation}\delta=\frac{\sin \left(\frac{F}{D} \sqrt{x^{2}+y^{2}+\frac{F}{2}}\right)}{50 F} \frac{1}{\left.1+e^{\left[-0.6\left(0.4 D-\sqrt{x^{2}+y^{2}}\right)\right]}\right.},\end{equation}
where the focal length $F=33\mathrm{~m}$, the diameter of antenna $D=110\mathrm{~m}$.

We found that the first-order and second-order partial derivatives of $\delta$ are all about $10^{-4} \mathrm{~m}$ and the product of any two of them is about  $10^{-7} \mathrm{~m}$ to $10^{-8} \mathrm{~m}$, as shown in Fig~\ref{fig3b} and Fig~\ref{fig3c}. Thus all the product-terms seem negligible for $a_{11}$ and $a_{22}$.

Substituting Eq.(3) and Eq.(5) into Eq.(7), $a_{11}$ and $a_{22}$ can be solved below.
\begin{equation}
\begin{split}
\begin{aligned}
a_{11}&=\frac{f_{r r} r^{3}+f_{\varphi}^{2} f_{r r} r-f_{\varphi} f_{r} f_{r \varphi} r+f_{\varphi}^{2} f_{r}}{\left[r^{2}\left(1+f_{r}^{2}\right)+f_{\varphi}^{2}\right]^{\frac{3}{2}}}\\
&=\frac{-\frac{\delta_{\theta} \delta_{r \varphi}}{2 F} r^{2}-\delta_{r} \delta_{r \varphi} \delta_{\varphi} r+\frac{\delta_{\varphi}^{2} r}{F}+\delta_{r r} \delta_{\varphi}^{2} r+\frac{r^{3}}{2 F}+\delta_{r r} r^{3}+\delta_{r} \delta_{\varphi}^{2}}{\left(\frac{r^{4}}{4 F^{2}}+\frac{r^{3} \delta_{r}}{F}+r^{2} \delta_{r}^{2}+r^{2}+\delta_{\varphi}^{2}\right)^{\frac{3}{2}}}\\
&\approx \frac{\frac{r^{3}}{2 F}+\delta_{r r} r^{3}}{r^{3}\left(1+\frac{r^{2}}{4 F^{2}}\right)^{\frac{3}{2}}} =\frac{\frac{1}{2 F}+\delta_{r r}}{\left(1+\frac{r^{2}}{4 F^{2}}\right)^{\frac{3}{2}}}=-\frac{\frac{1}{2 F}+\delta_{r r}}{\left(1+\frac{z}{F}\right)^{\frac{3}{2}}}\\
a_{22}&=\frac{-r\left(f_{r}^{2}+1\right) f_{\varphi \varphi}-f_{r}\left[r^{2}\left(f_{r}^{2}+1\right)-r f_{\varphi} f_{\varphi r}+f_{\varphi}^{2}\right]}{\left[r^{2}\left(1+f_{r}^{2}\right)+f_{\varphi}^{2}\right]^{\frac{3}{2}}}
 \approx \frac{\left(\frac{z}{F}+1\right) \delta_{\varphi \varphi}+\frac{r^{4}}{8 F^{3}}+\frac{r^{2}}{2 F}}{r^{2}\left(1+\frac{z}{F}\right)^{\frac{3}{2}}}.
\end{aligned}
\end{split}
\end{equation}

\subsection{Phase Matching} 

In the previous section the curvature matrix of the deformed antenna surface is obtained through the Weingarten maps. The detection method studied in this paper is that the plane wave incident on the antenna reflector, and the reflected electromagnetic wave radiates to the hemispherical detection field with the focus of the rotating paraboloid, namely, the feed position, as the spherical center.

   \begin{figure}[H]
   \centering
   \includegraphics[width=0.7\textwidth, angle=0]{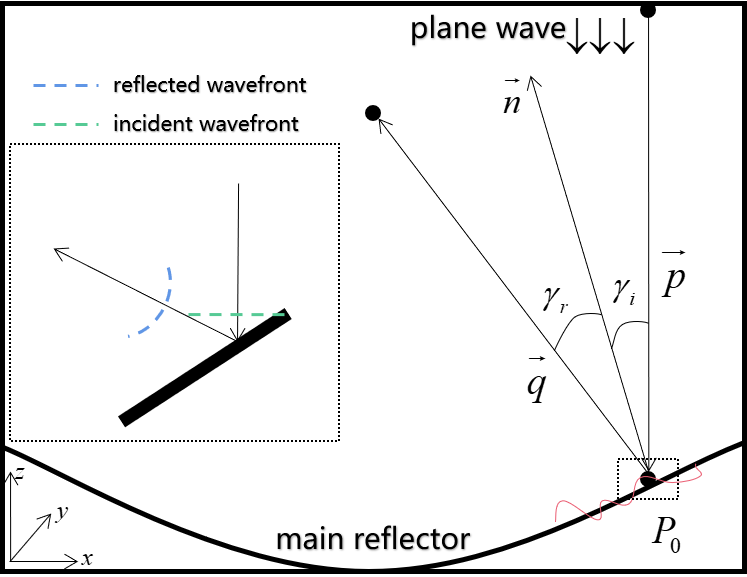}
   \caption{Schematic diagram of phase matching}
   \label{fig4}
   \end{figure}

Therefore, to study the electric field intensity of hemispherical detection field, we need to focus on the reflection wavefront of plane wave reflected by antenna reflector according to the principle of geometrical optics. This section mainly discusses the relation between the curvature matrix of incident wavefront, reflected wavefront and reflection surface.

The phase matching relation of incident and reflection wavefront on convex surface is given below based on GO theory(\citealt{Lee+1974})
\begin{equation}[a] \cos \gamma_{i}+K_{\mathrm{i}}^{T} Q^{\mathrm{i}}\left(P_{0}\right) K_{\mathrm{i}}=-[a] \cos \gamma_{\mathrm{r}}+K_{\mathrm{r}}^{T} Q^{\mathrm{r}}\left(P_{0}\right) K_{\mathrm{r}},\end{equation}
where $[a]$  is the curvature matrix of the reflector surface, and $\gamma _{i}$ is the incident angle while $\gamma _{r}$ is the reflect angle. $Q^{\mathrm{i}}\left(P_{0}\right)$ is the curvature matrix of the incident wavefront, and $Q^{\mathrm{r}}\left(P_{0}\right)$ is the curvature matrix of the reflect wavefront, as shown in Fig~\ref{fig4}.

$K_{i}$ and $K_{r}$ are transformation matrices of orthogonal basis, as shown below. $K_{i}^{T}$ and  $K_{r}^{T}$  are the transposes of $K_{i}$ and $K_{r}$.
\begin{equation}K_{\mathrm{i}}=\left(\begin{array}{cc}1 & 0 \\ 0 & -\cos \gamma_{\mathrm{i}}\end{array}\right), \quad K_{\mathrm{r}}=\left(\begin{array}{cc}1 & 0 \\ 0 & \cos \gamma_{\mathrm{r}}\end{array}\right).\end{equation}

Supposed the incident wavefront is a plane,$Q^{\mathrm{i}}\left(P_{0}\right)=0$ , and according to Fermat principle, we have $\gamma _{i}=\gamma _{r}$ .Thus, the curvature matrix of the reflect wavefront  $Q^{\mathrm{r}}\left(P_{0}\right)$ can be expressed as Eq.(12). (The negative sign comes from the concave surface.)
\begin{equation}Q^{r}=-2 \cos \left(\gamma_{r}\right)\left(K_{r}^{T}\right)^{-1}[a] K_{r}^{-1}=\left(\begin{array}{ll}Q_{11} & Q_{12} \\ Q_{21} & Q_{22}\end{array}\right),\end{equation}
where
\begin{equation}Q_{11}=-\frac{2}{\cos \gamma} a_{11}, \quad Q_{12}=2 a_{12},\quad Q_{21}=2 a_{21}, \quad Q_{22}=-2 \cos \gamma a_{22}.\end{equation}
We have also
\begin{equation}\cos (\gamma)=\left\langle n,[0,0,1]^{T}\right\rangle=\frac{r}{\sqrt{r^{2}\left(1+f_{r}^{2}\right)+f_{\varphi}^{2}}} \approx \frac{1}{\left(\frac{z}{F}+1\right)^{\frac{1}{2}}}.\end{equation}

From the conclusion of differential geometry, principal curvatures of reflected wavefront $\frac{1}{R_{1}}$ , $\frac{1}{R_{2}}$ are eigenvalues of curvature matrix(\citealt{Lee+etal+1979})
\begin{equation}\frac{1}{R_{1}}, \frac{1}{R_{2}}=\lambda_{1}, \lambda_{2}=\frac{1}{2}\left[\left(Q_{11}+Q_{22}\right) \pm \sqrt{\left(Q_{11}-Q_{22}\right)^{2}+4 Q_{12} Q_{21}}\right].\end{equation}

\begin{figure}[H]
   \centering
   \includegraphics[width=0.7\textwidth, angle=0]{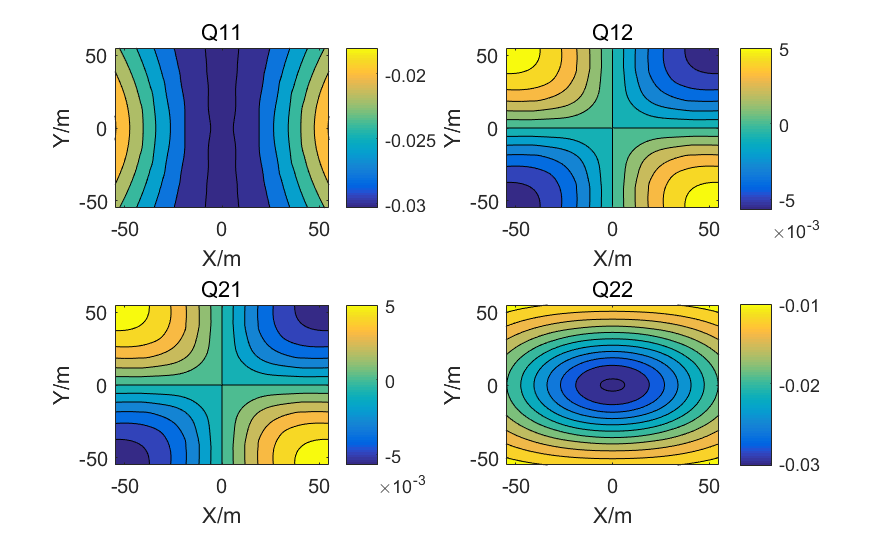}
   \caption{Magnitude of $Q_{11}$,$Q_{12}$,$Q_{21}$ and $Q_{22}$ of deformed antenna surface calculated by MATLAB, respectively. Where the deformation is assumed as Eq.(8) and generally the effect of small deformation on the magnitudes above seems to be negligible}
   \label{fig5}
   \end{figure}

As shown in Fig~\ref{fig5}, for the deformed reflector $f (x, y)$, the magnitude of $Q_{11}$ and $Q_{22}$ calculated in MATLAB are in the order of  $10^{-2} \mathrm{~m}$, while  $Q_{12}$ and  $Q_{21}$ are about  $10^{-3} \mathrm{~m}$. Additionally, the larger absolute values are outside the range of antenna aperture, and the absolute value near the aperture center is close to $0$. Therefore, to facilitate subsequent derivations,$Q_{12}$ and  $Q_{21}$ in Eq.(15) are ignored.

Thus, according to Eq.(14), the two principal curvatures of the reflected wavefront $1/{R_{1}},1/{R_{2}}$ are equal to $Q_{11}$ and $Q_{22}$, respectively. Put Eq.(9) and Eq.(14) into Eq.(13), $Q_{11}$ can be obtained as
\begin{equation}Q_{11}=-\frac{2}{\cos \gamma} \frac{\frac{1}{2 F}+\delta_{r r}}{\left(1+\frac{z}{F}\right)^{\frac{3}{2}}}=-\frac{1+2 F \delta_{r r}}{F+z}.\end{equation}
We have also
\begin{equation}Q_{22}=-2 \cos \gamma a_{22}=-\frac{2 F \delta_{\varphi \varphi}+r^{2}}{r^{2}(F+z)}.\end{equation}

\subsection{Deformation-intensity Equation}

\begin{figure}[H]
   \centering
   \includegraphics[width=0.4\textwidth, angle=0]{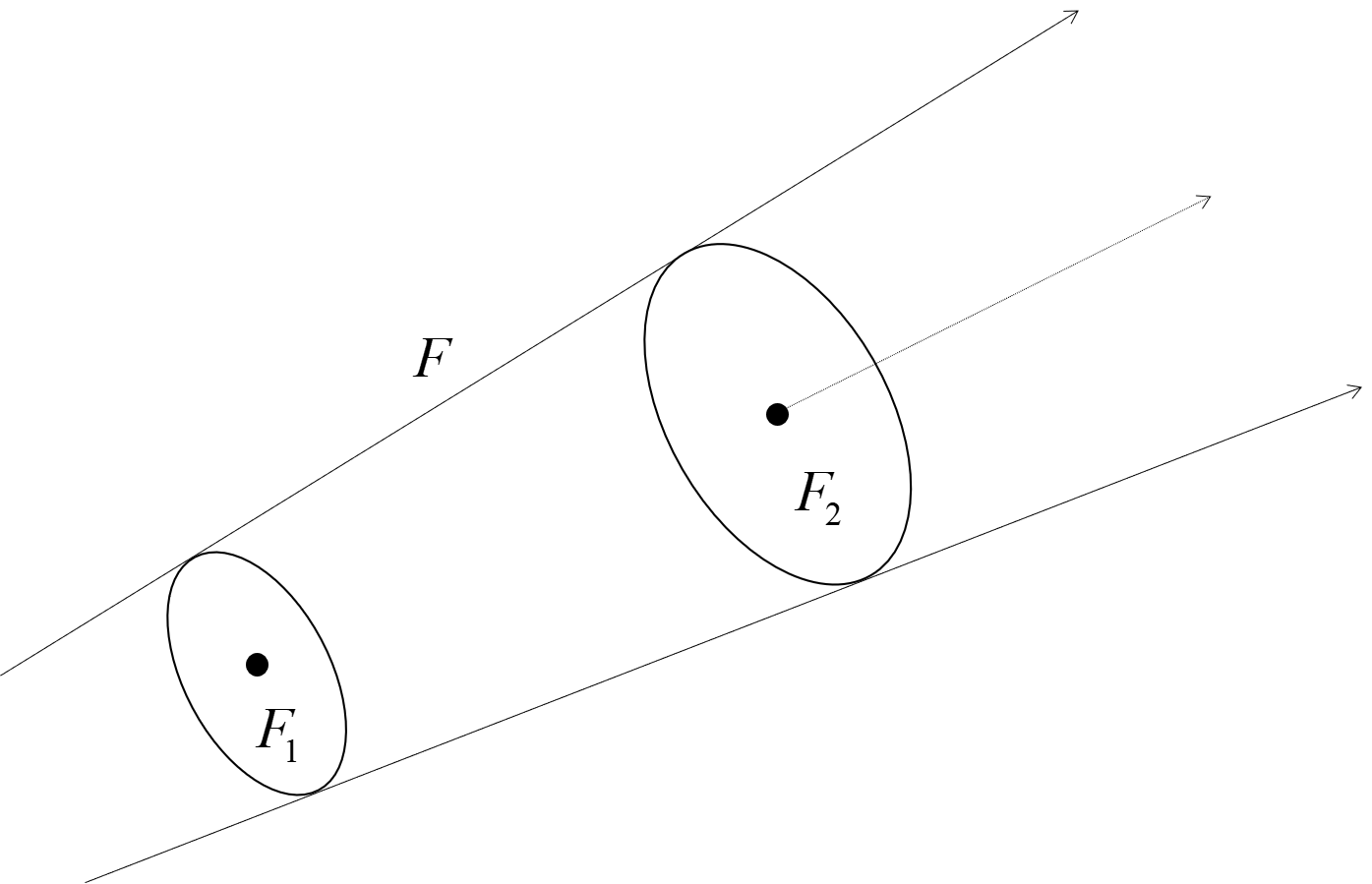}
   \caption{Lights travel in a homogeneous medium within a ray tube}
   \label{fig6}
   \end{figure}

According to the intensity law of geometrical optics, namely, the total energy flow of light along a ray tube is constant(\citealt{Deschamps+1972}), as shown in Eq.(18).
\begin{equation}\int_{F_{1}} \vec{W} \cdot d \alpha=-\int_{F_{2}} \vec{W} \cdot d \alpha,\end{equation}
where  $F_{1}$ and $F_{2}$ are the cross-sectional areas of the tube, respectively. $\vec{W}=1 / 2 \vec{E} \times \vec{H}$  is the energy flow density, namely, Poynting vector(\citealt{Barbaraci+2020}), whose direction is parallel to the surface $F$ of the ray tube, as shown in Fig~\ref{fig6}.

Then the expression of the field propagating along the geometrical optics ray in a homogeneous medium is obtained as follows.
\begin{equation}E_{2}=E_{1} \sqrt{\frac{\rho_{1} \rho_{2}}{\left(\rho_{1}+s\right)\left(\rho_{2}+s\right)}} e^{-j k s},\end{equation}
where  $E_{1}(V/m)$ is the electric field at $s=0$. $ e^{-j k s}$ is spatial phase delay factor. $\rho_{1}$ and $\rho_{2}$ (m) are the two principal curvatures of the wavefront. And $k=\omega / v=2 \pi / \lambda$ is propagation constant.

\begin{figure}[htbp]
\centering
\subfloat[]{
\label{fig7a}
  \begin{minipage}[t]{0.45\textwidth}
  \centering
  \includegraphics[width=2in]{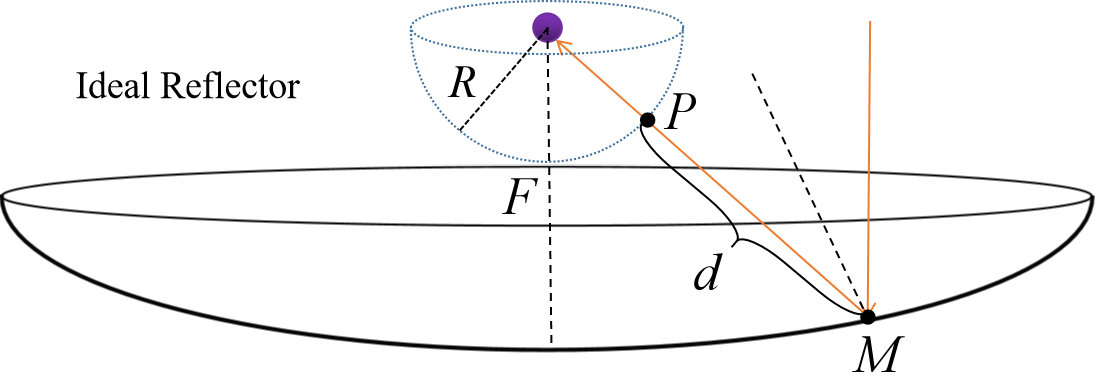}
  \end{minipage}
}
\subfloat[]{
\label{fig7b}
  \begin{minipage}[t]{0.45\textwidth}
  \centering
  \includegraphics[width=2in]{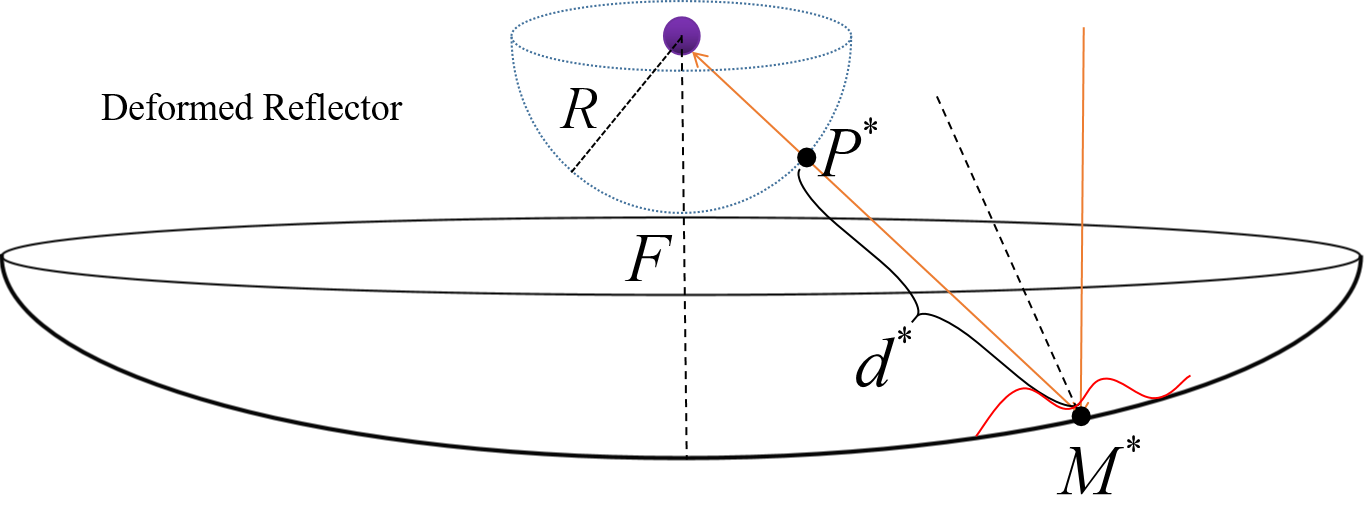}
  \end{minipage}
}
\caption{The reflection process of the Part(a) and part(b) describe the reflection process of ideal surface and deformed surface, respectively.}
\end{figure}

Then, for the ideal antenna surface, as shown in Fig~\ref{fig7a} above. Suppose a reflection point on the antenna reflector is $M$, and the incident plane wave is reflected by the reflector and propagates to point $P$ of the spherical field with a distance of $d$, which can be obtained by $d=F+z-R$ , from point $M$. The electric field at point $P$ is given by
\begin{equation}E_{P}=E_{M} \sqrt{\frac{R_{1} R_{2}}{\left(R_{1}+d\right)\left(R_{2}+d\right)}} e^{-j k d}.\end{equation}

Now put the expression of ideal reflector $f=z=r^{2} / 4 F$  into Eq.(16) and Eq.(17), the two principal curvatures of the reflected wavefront can be obtained as
\begin{equation}\left.Q_{11}\right|_{\delta=0}=\left.Q_{22}\right|_{\delta=0}=-\frac{1}{F+z}.\end{equation}

Put Eq.(21) into Eq.(20), the amplitude of electric field at point $P$ in ideal reflector can be expressed as below.
\begin{equation}\left|E_{P}\right|=\left|E_{M}\right| \frac{F+z}{R}.\end{equation}

Considering the reflector with deformation as shown in Fig~\ref{fig7b}, we suppose a reflection point on the deformed surface is $M^{*}$, and then the wave reflected by the deformed reflector propagates to the point$P^{*}$ with a distance of $d^{*}$ from $M^{*}$.

For a large-scale rotating parabolic antenna, generally, the deformation of antenna surface is less than $10^{-3}$ order of magnitude, which is very small relative to its radius. Thus two assumptions can be made below:

(1)$d^{*}=\left|P^{*} M^{*}\right| \approx|P M|=d=F+z-R.$

(2)The point $P^{*}$ can be approximately regarded as $P$.

Based on the assumptions above, the amplitude at point $P^{*}$ on the spherical field can be given by
\begin{equation}\left|E_{P} *\right|=\left|E_{M}\right| \sqrt{\frac{R_{1} R_{2}}{\left(R_{1}+d\right)\left(R_{2}+d\right)}}=\left|E_{M}\right| \sqrt{\frac{1}{\left(1+Q_{11} d\right)\left(1+Q_{22} d\right)}}.\end{equation}

Compare Eq.(23) with Eq.(22), we can get
\begin{equation}\left|\frac{E_{P}}{E_{P}^{*}}\right|^{2}=\left(\frac{F+z}{R}\right)^{2}\left(1+Q_{11} d\right)\left(1+Q_{22} d\right).\end{equation}

Then put Eq.(16) and Eq.(17) into Eq.(24), and we also regard the product-terms of partial derivatives of $\delta$ much smaller.
\begin{equation}
\begin{split}
\begin{aligned}
\left|\frac{E_{P}}{E_{P} *}\right|^{2}&=\left(\frac{F+z}{R}\right)^{2}\left(1-\frac{1+2 F \delta_{r r}}{F+z} d\right)\left(1-\frac{2 F \delta_{\varphi \varphi}+r^{2}}{r^{2}(F+z)} d\right)\\
&=\left(\frac{F+z}{R}\right)^{2}\left[1-\frac{2 r^{2}+2 F\left(r^{2} \delta_{r r}+\delta_{\varphi \varphi}\right)}{r^{2}(F+z)} d+\frac{r^{2}+2 F\left(r^{2} \delta_{r r}+\delta_{\varphi \varphi}\right)+4 \mathrm{~F}^{2} \delta_{r r} \delta_{\varphi \varphi}}{r^{2}(F+z)^{2}} d^{2}\right]\\
&\approx\left(\frac{F+z}{R}\right)^{2}-\frac{2 r^{2}+2 F\left(r^{2} \delta_{r r}+\delta_{\varphi \varphi}\right)}{r^{2} R^{2}} d(F+z)+\frac{r^{2}+2 F\left(r^{2} \delta_{r r}+\delta_{\varphi \varphi}\right)}{r^{2} R^{2}} d^{2}.
\end{aligned}
\end{split}
\end{equation}

A transformation from cylindrical coordinates to Cartesian coordinates, as shown in Eq.(26) and Eq.(27), is introduced.
\begin{equation}
\delta_{r r}=\delta_{x x} \frac{x^{2}}{x^{2}+y^{2}}+2 \delta_{x y} \frac{x y}{x^{2}+y^{2}}+\delta_{y y} \frac{y^{2}}{x^{2}+y^{2}}.
\end{equation}
\begin{equation}
\delta_{\varphi \varphi}=y^{2} \delta_{x x}-2 x y \delta_{x y}+x^{2} \delta_{y y}-x \delta_{x}-y \delta_{y}.
\end{equation}

Then, put Eq.(26) and Eq.(27) into Eq.(25).
\begin{equation}
\begin{split}
\begin{aligned}
\left|\frac{E_{P}}{E_{P} *}\right|^{2}
&=\left(\frac{d+R}{R}\right)^{2}-\frac{2 r^{2}+2 F r^{2}\left(\delta_{x x}+\delta_{y y}\right)}{r^{2} R^{2}} d(d+R)+\frac{r^{2}+2 F r^{2}\left(\delta_{x x}+\delta_{y y}\right)}{r^{2} R^{2}} d^{2} \\
&=\frac{d^{2}+2 R d+R^{2}}{R^{2}}-\frac{d^{2}}{R^{2}}-\frac{2 d+2 F \Delta \delta d}{R} \\
&=1-\frac{2 F d}{R} \Delta \delta.
\end{aligned}
\end{split}
\end{equation}

Where $d=F+z-R$. We finally obtain the deformation-intensity equation of the spherical field around the feed as Eq.(28).

\subsection{Numerical Algorithm}

According to Eq.(28), the deformation-intensity equation can be written as
\begin{equation}\Delta \delta=F A G,\end{equation}
where $FAG=FA/G, G=-2Fd/R, FA=|E_{P}|^{2}/|E_{P}^{*}|^{2}-1=EA-1$.

As shown in Eq.(29), the deformation-intensity equation of spherical field is finally transformed into the form of a standard Poisson equation, where $FA$ and $G$ are  discrete matrices relative to aperture coordinates $(x, y)$. Thus we can only find the numerical solution of the Poisson equation above, which is a typical elliptic equation that can be solved precisely by a traditional method, finite difference method(FDM).

The main idea of FDM is to transform the Laplacian for the calculation of $\Delta\delta$ into a discrete scheme. Typically, the third-order Laplacian can be given by a five point difference scheme as shown in Eq.(30) based on Taylor expansion(\citealt{Sauer+2014}).
\begin{equation}L=\left(\begin{array}{ccc}0 & 1 & 0 \\ 1 & -4 & 1 \\ 0 & 1 & 0\end{array}\right).\end{equation}

Thus the matrix $FAG$ in Eq.(29) can be expressed as the convolution of $L$ to the deformation $\delta$.
\begin{equation}FAG = \delta * L .\end{equation}

By means of $FFT$, $\delta$ in Eq.(31) can be solved efficiently in frequency domain. However, the accuracy of results is inadequate for the demand of surface deformation recovery. In order to apply the FDM, discrete Laplacian $L$ in Eq.(30) has to be transformed into matrix $T$, as shown in Eq.(32).
\begin{equation}T=\frac{1}{h^{2}}\left(\begin{array}{ccccc}T_{0} & -I & 0 & \cdots & 0 \\ -I & T_{0} & \ddots & \ddots & \vdots \\ 0 & \ddots & \ddots & \ddots & 0 \\ \vdots & \ddots & \ddots & T_{0} & -I \\ 0 & \cdots & 0 & -I & T_{0}\end{array}\right)_{N^{2} \times N^{2}},\end{equation}
where
$$T_{0}=\left(\begin{array}{ccccc}4 & -1 & 0 & \cdots & 0 \\ -1 & 4 & \ddots & \ddots & \vdots \\ 0 & \ddots & \ddots & \ddots & 0 \\ \vdots & \ddots & \ddots & 4 & -1 \\ 0 & \cdots & 0 & -1 & 4\end{array}\right)_{N \times N},$$
and h is step length. Thus we have
\begin{equation}F A G_{N^{2} \times 1}=T_{N^{2} \times N^{2}} \times \delta_{N^{2} \times 1}.\end{equation}

Finally, the deformation has been converted to the solution of a linear equation, which is then straightforward to employ many iterative methods such as Jacobian method, Gauss-Seidel method and successive over-relaxation(SOR) method(\citealt{Sauer+2014}). In our experiments we have employed Gauss-Seidel method that has demonstrated excellent performance. In fact, comparing with FFT and left division methods, we have proved it again.

\section{SIMULATION}
\subsection{GO Method}

We have determined the effectiveness of the numerical algorithm by simulation. The simulation conditions are shown in Table~\ref{Tab2}.
\begin{table}[H]
\begin{center}
\caption[]{ Simulation conditions}\label{Tab2}

 \begin{tabular}{cccc}
  \hline\noalign{\smallskip}
Diameter of the antenna &	Focal length  &	 Radius of spherical field	&  Frequency\\
$D/\mathrm{~m}$ &	$F/\mathrm{~m}$	& $R/\mathrm{~m}$ & $f/\mathrm{~GHZ}$\\
$110$	& $33$ & $10$ & $0.3$ \\
  \hline\noalign{\smallskip}
Source & Taper angle & Number of field points($\theta * \phi$) & Magnitude of $\delta$ ($M_{\delta}/\mathrm{~mm}$)   \\ 
plane wave & $79.61^{\circ}$ & $90*360$ & $1.2$ \\

  \noalign{\smallskip}\hline
\end{tabular}
\end{center}
\end{table}
Primary steps are listed below.

\begin{figure}[htbp]
\centering
\subfloat[]{
\label{fig8a}
  \begin{minipage}[t]{0.45\textwidth}
  \centering
  \includegraphics[width=2.5in]{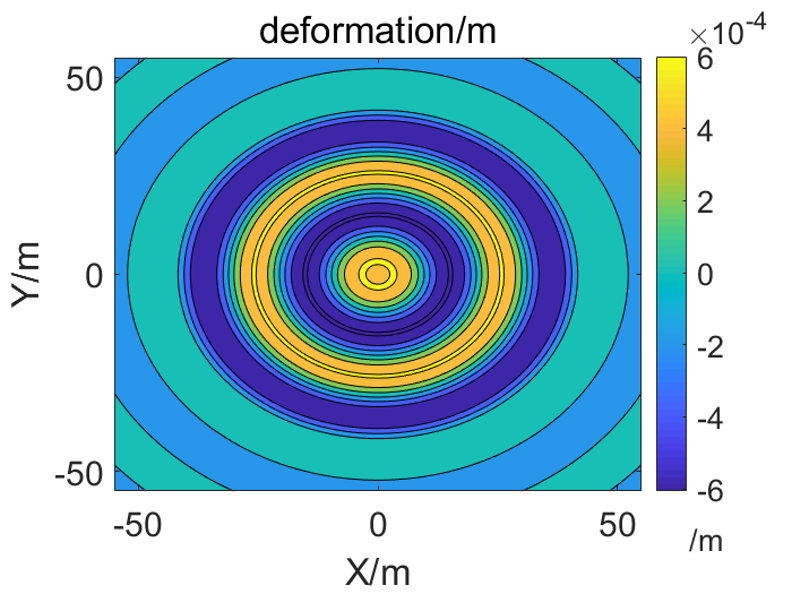}
  \end{minipage}
}
\subfloat[]{
\label{fig8b}
  \begin{minipage}[t]{0.45\textwidth}
  \centering
  \includegraphics[width=2.5in]{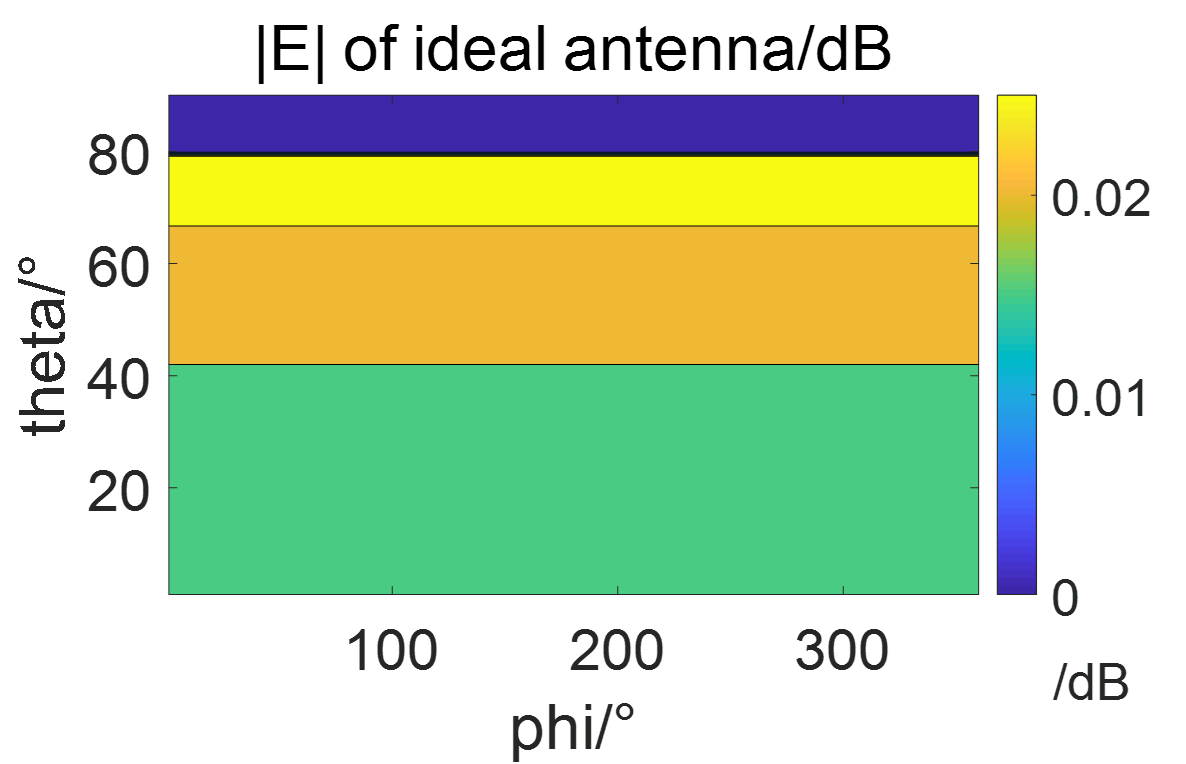}
  \end{minipage}
}

\subfloat[]{
\label{fig8c}
  \begin{minipage}[t]{0.45\textwidth}
  \centering
  \includegraphics[width=2.5in]{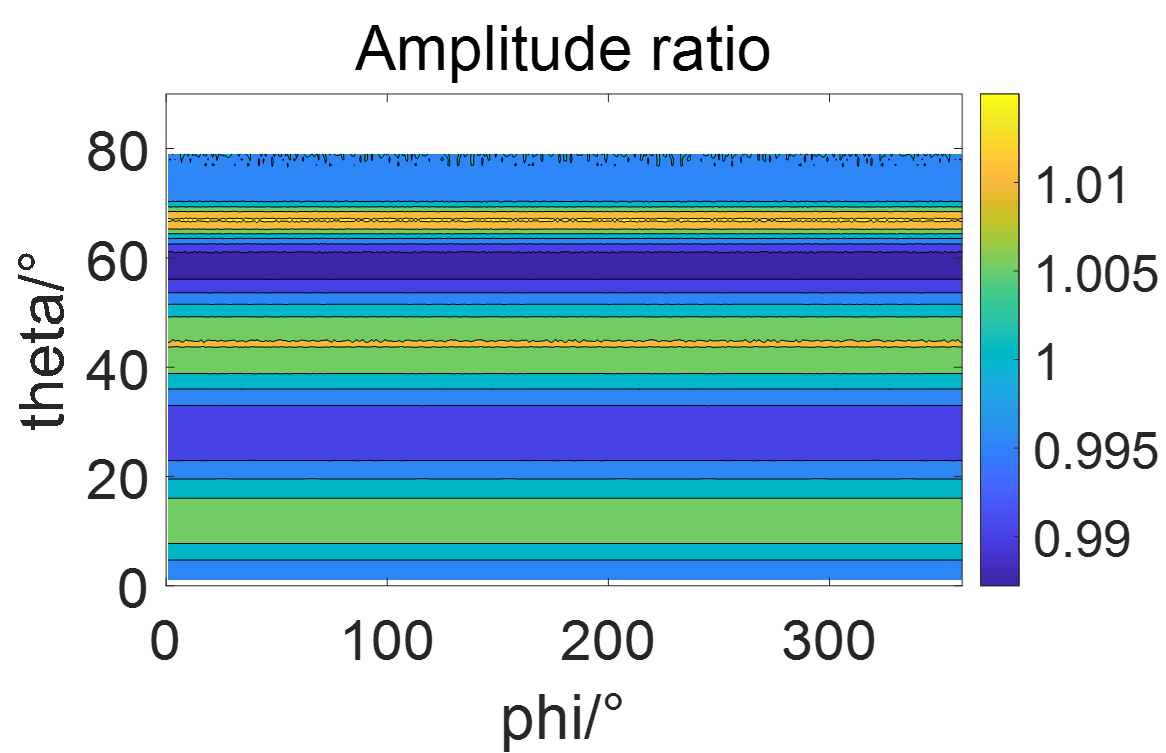}
  \end{minipage}
}
\subfloat[]{
\label{fig8d}
  \begin{minipage}[t]{0.45\textwidth}
  \centering
  \includegraphics[width=2.5in]{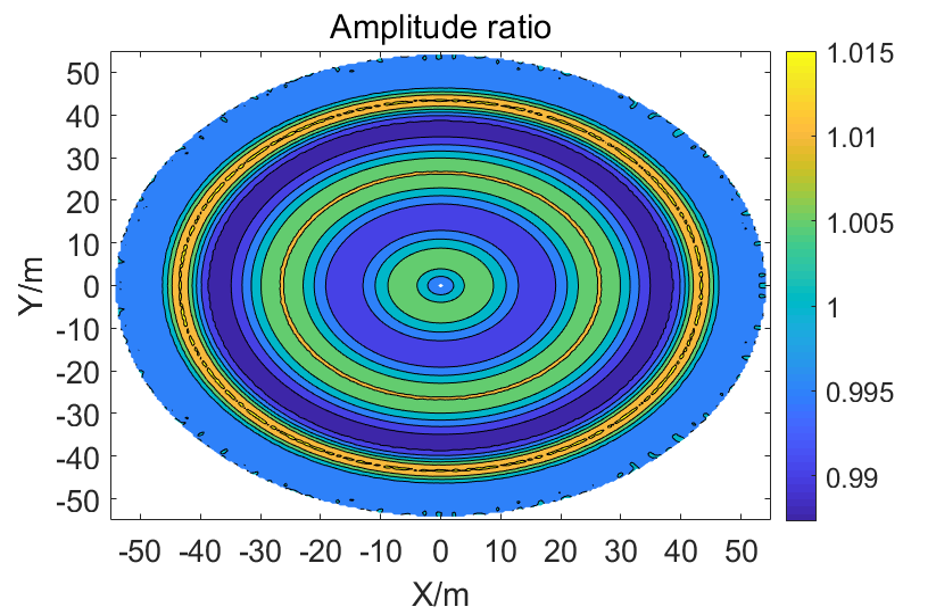}
  \end{minipage}
}
\caption{The simulation results of hemispherical filed by GO method. (a) and (b) give the assumed deformation and simulated amplitudes in case of ideal surface. (c) and (d) describe the FA before and after interpolation, respectively.}
\end{figure}

(1) Design the global smooth deformation $\delta$ as Eq.(8) and Fig~\ref{fig8a}.          

(2) Generate the amplitudes $|E|$ and $|E^{*}|$ of proposed spherical field with and without deformation on the main reflector by GO method, respectively. The simulated amplitude distribution of ideal antenna,$|E|$, is shown as Fig~\ref{fig8b}. The intensity ratio in Eq.(28), $|E|^{2}/|E^{*}|^{2}$, is given as Fig~\ref{fig8c}.

(3) Interpolate the simulated data $EA(\theta, \varphi)$ into Cartesian coordinate of aperture, namely,$EA^{\prime}(x, y)$. According to the antenna geometry as a rotating paraboloid, the relations between ($\theta$, $\phi$) and ($x, y$) are given as
\begin{equation}\theta=\arccos \left(\frac{4 F^{2}-x^{2}-y^{2}}{4 F^{2}+x^{2}+y^{2}}\right), \varphi=\left\{\begin{array}{l}\arctan \left(\frac{y}{x}\right), x>0, y>0 \\ \arctan \left(\frac{y}{x}\right)+\pi, x<0 \\ \arctan \left(\frac{y}{x}\right)+2 \pi, x>0, y<0\end{array}\right..\end{equation}

\begin{figure}[H]
   \centering
   \includegraphics[width=\textwidth, angle=0]{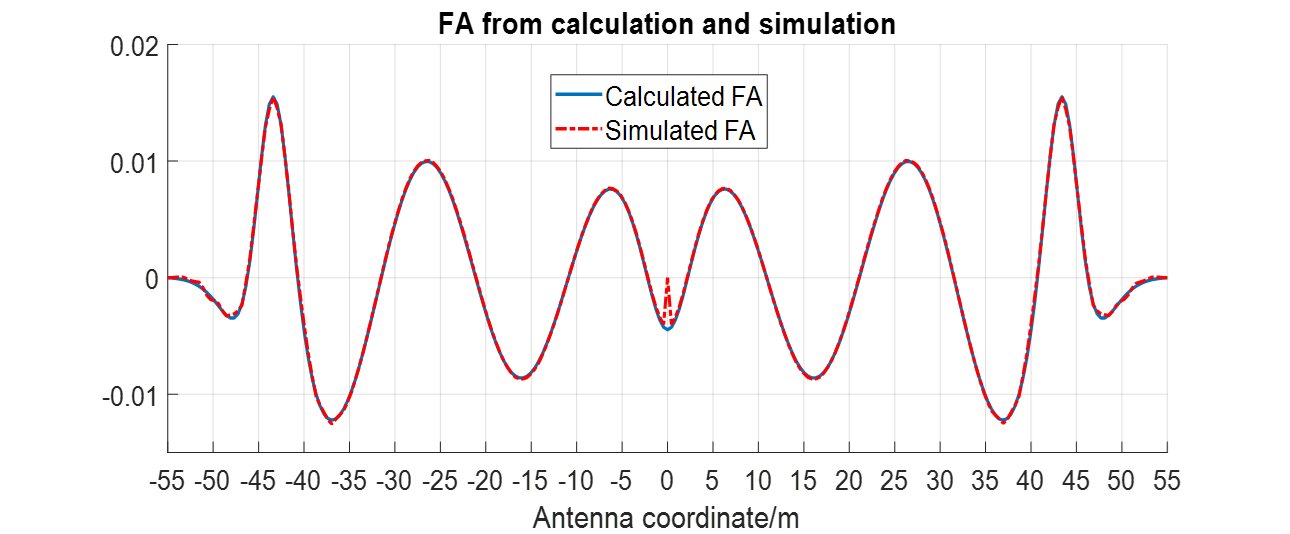}
   \caption{Comparison of intensity ratio calculated from Eq.(28) with simulation results by GO method}
   \label{fig9}
   \end{figure}

Then $EA$ can be shown as Fig~\ref{fig8d}. Furthermore, we compared the simulated $FA$ with the calculated value $G \cdot \nabla^{2} \delta$ . By GO method, it was practically coincided with the theoretical values, as shown in Fig~\ref{fig9}, which verified the effectiveness of the deformation-intensity equation we proposed above.

(4) Recover $\delta$ from $FA$ by the proposed algorithm and compare the results $\delta_{r}$ with the ideal deformation $\delta$. In order to make a quantitative comparison, root-mean-square error (RMS) and relative root-mean-square error (RRMS) are introduced as below(\citealt{Huang+etal+2017}).
\begin{equation}R M S=\sqrt{\frac{\sum_{m} \sum_{n}\left[\delta_{r}(m, n)-\delta(m, n)\right]^{2}}{N^{2}}},\end{equation}
\begin{equation}R R M S=R M S / M_{\delta}.\end{equation}

\begin{figure}[H]
   \centering
   \includegraphics[width=\textwidth, angle=0]{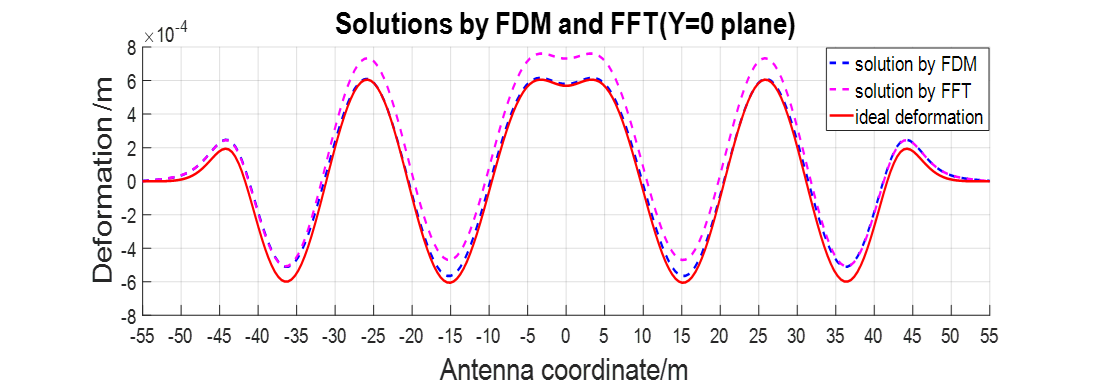}
   \caption{Comparison of solutions by FDM and FFT with the assumed deformation according to the results of GO simulation.}
   \label{fig10}
   \end{figure}

The inference proposed above was verified by a numerical calculation. As is shown in Fig~\ref{fig10}, the solution of deformation by FDM has an accuracy of $RRMS= 4.06\%$. As a comparison, the $RRMS$ of solution by FFT is $7.47\%$.

\subsection{PO\&PTD Method}

According to the Maxwell equations, the well-known physical optic method (PO) can be employed to calculate the electromagnetic field reflected by antenna based on the surface current integration, precisely(\citealt{Umul+2020}). Compared with GO method, which ignores the volatility of light, PO method is widely regarded as a more accurate theory when calculating the reflected field of antenna radiation. Furthermore, physical theory of diffraction(PTD) can be a supporting method to calculate the diffraction field caused by the reflector rim in simulation.

\begin{figure}[H]
   \centering
   \includegraphics[width=\textwidth, angle=0]{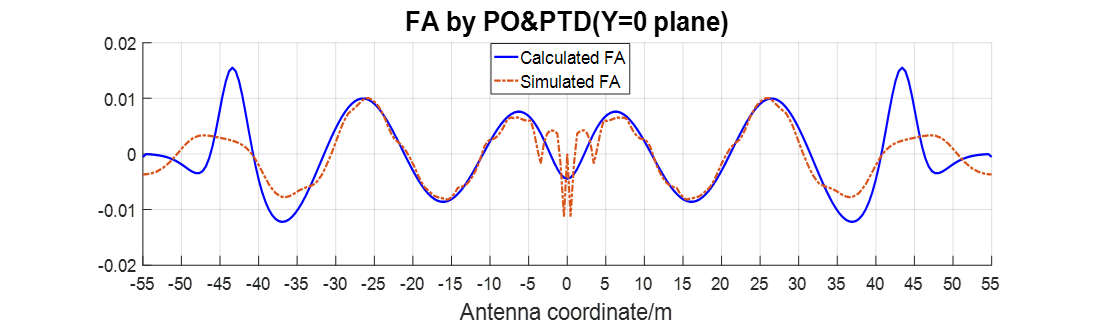}
   \caption{Comparison of intensity ratio calculated from Eq.(28) with simulation results by PO\&PTD}
   \label{fig11}
   \end{figure}

Firstly, we keep the simulation condition unchanged as shown in Table1 and suppose the same deformation as Eq.(8). $FA$ simulated by PO\&PTD method is given as Fig~\ref{fig11}, which has a relatively large difference with the ideal value according to the proposed equation, especially on the outer domain of the aperture.

Then, the results of recovered deformation by PO\&PTD are given as following Fig~\ref{fig12}, the $RRMS$ of which by FDM and FFT are $7.14\%$ and $11.15\%$.

   \begin{figure}[H]
   \centering
   \includegraphics[width=\textwidth, angle=0]{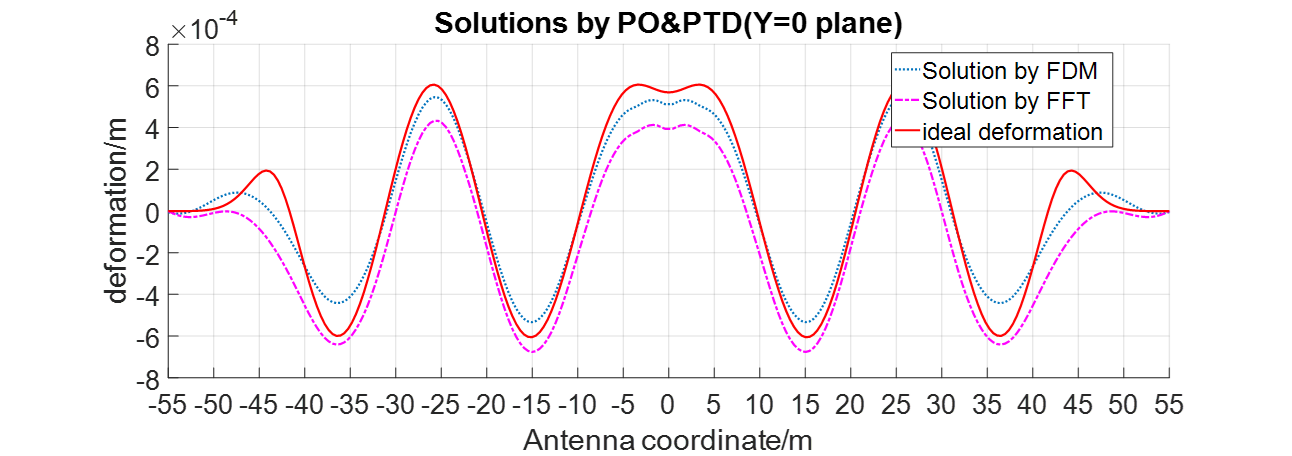}
   \caption{Comparison of solutions by FDM and FFT with the assumed deformation according to the results of PO\&PTD simulation.}
   \label{fig12}
   \end{figure}

Additionally, we have simulated the amplitudes of spherical field at different frequencies, and recovered the deformation by FDM, as shown in Table~\ref{Tab3}. Except for an invalid result at $0.1$GHZ, the $RRMS$ at different frequencies are mainly consistent. Especially, when the radiation frequency of plane wave is $0.6$GHZ or $1$GHZ, a more accurate result of deformation recovery is obtained.

\begin{table}[H]
\begin{center}
\caption[]{Results of algorithm in different frequencies by means of PO\&PTD}\label{Tab3}

 \begin{tabular}{cccccc}
  \hline\noalign{\smallskip}
Conditions & $RMS=327.65\mu m$ & $M_{\delta}=1.2mm$ & $R=10$ m & $N=256$ & $100\%$ region \\
algorithm & FDM & iterations & $1000$ & Time/s & $92 \pm 10$\\
  \hline\noalign{\smallskip}
$f$/GHZ & $0.1$ & $0.3$ & $0.6$ & $1$ & $3$\\
$RMS/\mu m$	& $264.84$ & $85.68$ &	$61.56$ &	$60.72$ & $77.04$\\
$RRMS/\%$ & $22.07$ &	$7.14$ & $5.13$ & $5.06$ & $6.42$\\
  \noalign{\smallskip}\hline
\end{tabular}
\end{center}
\end{table}

   \begin{figure}[H]
   \centering
   \includegraphics[width=\textwidth, angle=0]{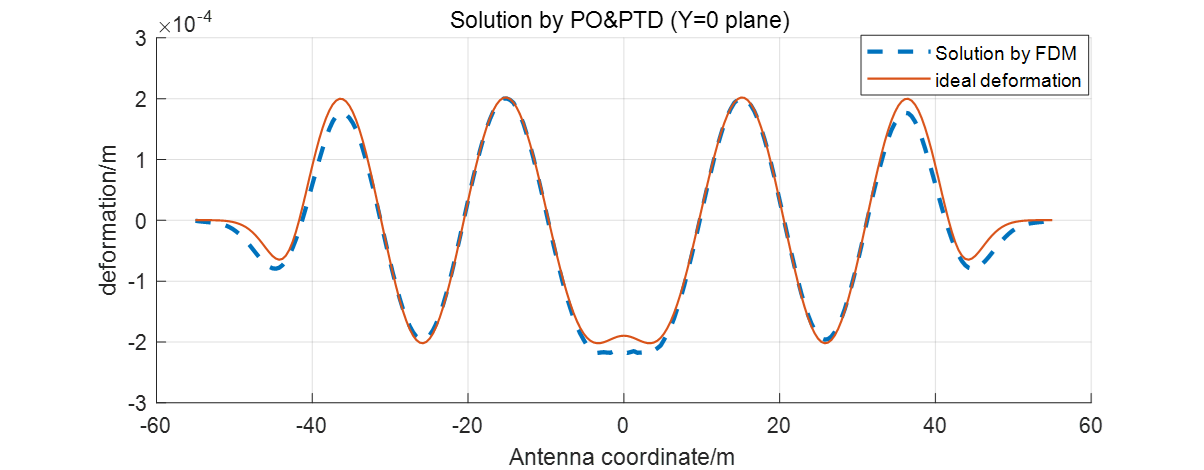}
   \caption{Solution by PO\&PTD method and FDM algorithm with the new assumed deformation}
   \label{fig13}
   \end{figure}
Moreover, we have changed a new deformation with a lower amplitude, whose $M_{\delta}$ is $0.4$mm, to further check the effectiveness of the measurement method presented in this paper. As is shown in Fig~\ref{fig13}, the red line is the deformation we set and the blue line is the solution solved by the FDM algorithm. Under this condition, the $RMS$ and $RRMS$ of the solution are respectively $16.95\mu$m and $4.20$\% while the $RMS$ of surface with the new deformation is about $109.22\mu$m.

\subsection{Errors Analysis}

(1) GO Limit

\begin{figure}[H]
   \centering
   \includegraphics[width=\textwidth, angle=0]{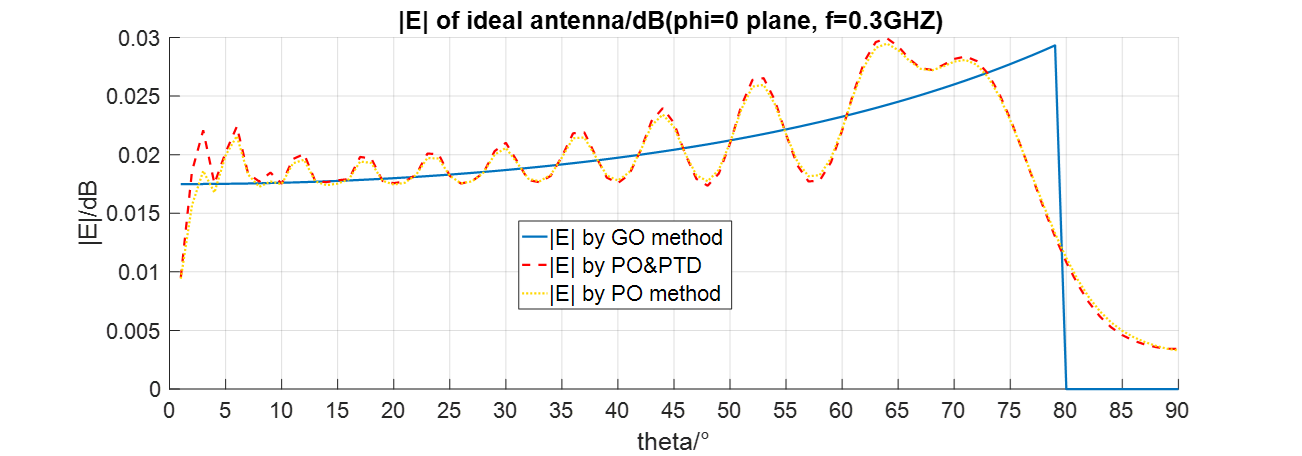}
   \caption{ Comparison of simulated amplitude distributions by GO, PO and PO\&PTD method}
   \label{fig14}
   \end{figure}

The basic errors in the derivation of the deformation-intensity equation is completely based on the theory of differential geometry and geometrical optics, and the wave character of electromagnetic wave is neglected in this process. However, the amplitude of spherical field by PO method is accurately calculated based on Maxwell equation, which leads to errors as shown in Fig~\ref{fig14}, and the errors will be larger with a lower frequency.

 As a result, there exist differences between the equation proposed above and the simulation results by PO method as shown in Fig~\ref{fig11}.

(2) Edge Diffraction

When the plane wave incident on the antenna reflector, diffraction occurs at the edge of the reflector, which will be calculated by PO\&PTD method in simulation. While the equation proposed fails to take it into account. As shown in Fig~\ref{fig14}, there exist subtle differences in simulation results of pure PO method (yellow line) and PO\&PTD (red line).

(3) Lack of Boundary Conditions

The process of solving a PDE precisely needs boundary conditions no matter what method is used(\citealt{Marinelli+etal+2021}). In fact, the boundary conditions of deformation on the antenna surface are usually unknown. Then a hypothetical initial value has to be used during solving the equation and the algorithm will adjust it step by step, which will lead to an unknown complex function included in the final solution.

\section{SUMMARY}

This paper presented a new method for deformation measurement of antenna main reflection surface based on geometric optics. Based on the Gaussian and Weingarten mapping in the differential geometry theory, the curvature matrix of any point on the deformed antenna surface was obtained. Then, with plane waves as sources radiating to the reflector, phase matching method was used to acquire the curvature matrix of the reflected electromagnetic wavefront. Furthermore, the relation between complex amplitudes of any two point on a ray can be given according to the theory of ray tube energy conservation in geometric optics. By this means, we derived the deformation-intensity equation thus to calculate the deformation on the antenna surface from the intensity of the hemispherical electromagnetic field with the feed as the sphere center. 

Despite making some approximations in the derivation process, the deformation -intensity equation proposed in this paper can provide good results, which has been verified through comparing with simulation results by PO method based on current integration. Additionally, the numerical algorithm based on FDM was employed to solve the PDE. The numerical results behaved more effective and stable compared with those by FFT.

To sum up, the main contributions of this study are as follows:

1. The method proposed in this paper effectively improves the measurement accuracy of the reflector surface, which can reach $16.95\mu\mathrm{m}$ while the original surface accuracy is about $109.22\mu\mathrm{m}$. As a comparison, under the same condition the errors of the phase coherent method, phase retrieval method and the near-field holography method are respectively about $25\mu\mathrm{m}$, $50\mu\mathrm{m}$ and $25\mu\mathrm{m}$. Moreover, the $RRMS$ has reduced by about $4$ percent compared with the near-field planar intensity scanning method under the same condition.

2. The measuring speed has been improved a lot. Firstly, compared with planar scanning, the scanning area of the hemisphere is only $1/15$ of the aperture and the total measuring time will be at least $10$ times less. Then, compared with the phase coherent holography method, there has no requirement to set another antenna and a dual channel reference receiver to calculate the reference phase, which will save a lot of measuring time.

3. The method proposed in this paper won’t be limited by the constraints of the type and frequency of the transmitter, the elevation angle and the measuring equipment. We can use not only the outer space satellite but also the far-field artificial beacon at any frequency to be the transmitter. And it is simple to select a hemisphere to be the scanning area whatever the elevation angle is so that we can realize the all-weather and all-attitude measurement for the deformation of the main reflector. Besides, there is no requirement for huge or complex equipment.

\section{ACKNOWLEDGMENT}

Supported by The National Key Research and Development Program of China of Research on Key Technologies of real-time shape control and ultra wideband pulsar signal processing for large aperture radio telescope.

Project U1931137 supported by National Natural Science Foundation of China.

\label{lastpage}

\end{document}